\documentclass[aps,superscriptaddress,nofootinbib,eqsecnum]{revtex4}

\usepackage{epsfig}

\input{epsf}
\begin{document}

\title{Introducing the notion of bare and effective mass via Newton's second law of motion.}

\author{Marcus Benghi Pinto}
\email{marcus@fsc.ufsc.br}
\affiliation{ Grupo de F\'{\i}sica Matem\'{a}tica e Teorias de Campos \\
Departamento de
F\'{\i}sica, Universidade Federal de Santa Catarina, 88040-900
Florian\'{o}polis, SC, Brazil}
\begin{abstract}
The concepts of bare and effective mass are widely used within modern physics.
Their meaning is discussed in advanced undergraduate and graduate courses such as solid state physics, nuclear physics and quantum field theory. Here I discuss how  these concepts may be introduced together with the discussion of Newton's second law of motion.  The setting up of  simple equations for the effective mass will allow instructors to discuss how external parameters, such as the temperature, influence this quantity. By expressing this type of equation as a power series one may also discuss perturbation theory and introduce  Feynman diagrams.
\end{abstract}

\maketitle

\section{Introduction}

Today, many undergraduate students majoring in physics attend introductory courses on elementary particle physics. This is only possible thanks to excellent textbooks such as the one by Griffiths \cite {griffiths}, who undertook the task of presenting this subject to students lacking a quantum field theory background. At the same time, this audience receives a lot of qualitative information concerning advanced subjects such as asymptotic freedom, phase transitions, symmetry breaking and restoration through serious popular books \cite {books},  articles \cite {sa} and the web \cite {sites}. Many times, these issues are discussed in conjunction with cosmological phase transitions, fascinating students.
 Very often, popular articles covering particle physics  refer to  quarks as having a bare mass as well as an effective mass \cite {books}. These topics also  appear in  advanced undergraduate courses, such as particle physics,  nuclear physics as well as solid state physics.

As far as elementary particle physics is concerned,  students will learn  that the $u$ and $d$ quarks are intrinsically very light, but within the confines of a hadron their effective mass is much greater \cite {griffiths}. In a typical solid state physics course it will be discussed that an electron, moving inside a solid, will respond in a different way to an externally applied  force than it would outside the solid. Therefore, due to the interactions of the electron with the lattice, its mass assumes different values inside  (effective mass) and outside  (bare mass) the solid \cite {cheng}. The same concepts can also appear in a nuclear physics course where the students will learn that the neutron mass inside a dense neutron star has a different value than it has at zero density.

These concepts may also be introduced in alternative ways, such as the one presented by
Mattuck \cite {mattuck}, where  Newton's second law of motion is used to define a classical quasi-particle propagator. However, although succeeding in giving the notion of bare and effective masses, this presentation is formulated in terms of a many-body system.  An excellent article, by Kittel \cite {kittel}, also treats the effective mass at the undergraduate level but it is mainly written to be used
in conjunction with a course in solid state physics.

Here I will present an alternative approach, which can be discussed   when one reviews Newton's laws of motion in a classical mechanics  course \cite {symon} for junior-level undergraduates.
Many of the students attending this type of course are familiar, at least from a qualitative point of view,   with terms such as effective mass and asymptotic freedom. In this work, I propose that these topics be discussed in a more formal way through  standard applications of  Newton's second law of motion.
It is notorious that students, eager to see modern physics in action,  do not see much fun in treating this classical problem and this paper has been written to allow the instructor, presenting the subject, to make a stimulating digression in order
to introduce  ideas currently used in modern physics. My aim is to set up, within the Newtonian formalism,   equations which will help undergraduate students to have a mathematical, rather than conceptual, view of non-trivial issues concerning modern theories. At the same time, this simple presentation could  be reviewed in more advanced courses, such as quantum field theory, as a preparation to the study of the renormalization programme. In this case, students will also profit from reading Refs. \cite {ren,hans} before tackling the subject.

In the next section, I will use Newton's second law of motion to define the  relation between bare mass and effective mass. I will discuss two cases starting with the notion of added mass, which occurs in fluid dynamics.  Next, I will consider   a block sliding  over a surface in the presence of friction to set up an equation which partially captures some features related to the concept of effective mass in modern theories. Contrary to the equation obtained in the fluid dynamics case, this  equation cannot be considered a true equation for the effective mass. Nevertheless, when used with care, it can be useful to discuss issues related to quantum field theories. In this same section, I will suppose that parameters such as the density and the coefficient of kinetic friction
 change with the temperature. This will allow us to
plot the effective mass in terms of the temperature, opening the possibility of discussing
issues related to chiral symmetry breaking/restoration in quantum chromodynamics (QCD), among others.

A short conceptual review of bare and effective masses in quantum field theories will be presented in section III.  The equation corresponding to the sliding block case will be rewritten as a series, in a diagrammatic way, in section IV. Finally, a brief comment on scale dependence which hints at the renormalization group equations is made in section V. The paper is written in a way that allows the instructor to adapt the material to be covered according to the background of her/his students. At a more introductory level, sections
II.A and III will suffice to introduce the notion of bare and effective masses. Sections II.B, IV, and V, can be left for students who are getting familiar with the language of quantum field theories. Section II.B must be presented with care to avoid confusions regarding Newton's second law of motion. To substantiate the discussion carried out  in section III, the students should be asked to read chapter 2 (especially sec. 2.3) of Ref. \cite {griffiths}, which gives a qualitative overview concerning issues such as asymptotic freedom. At the same time, the instructor should encourage more advanced students, such as the ones attending a course on particle physics, to perform some of the evaluations presented in Refs. \cite {ren,hans}.

\section{Bare and effective mass in classical mechanics}

In this section we will analyze Newton's second law of motion applied to two independent situations which will be useful to introduce ideas related to the concepts of bare mass and effective mass. In each case we shall see how an external paramater such as the temperature affects the effective mass. This procedure will allow a better understanding of the QCD chiral transition problem which will be briefly addressed in subsection III.C.

\subsection{ Added mass in fluid dynamics}
\begin{figure}[htb]
\vspace{0.5cm}
\centerline{\epsfig{figure=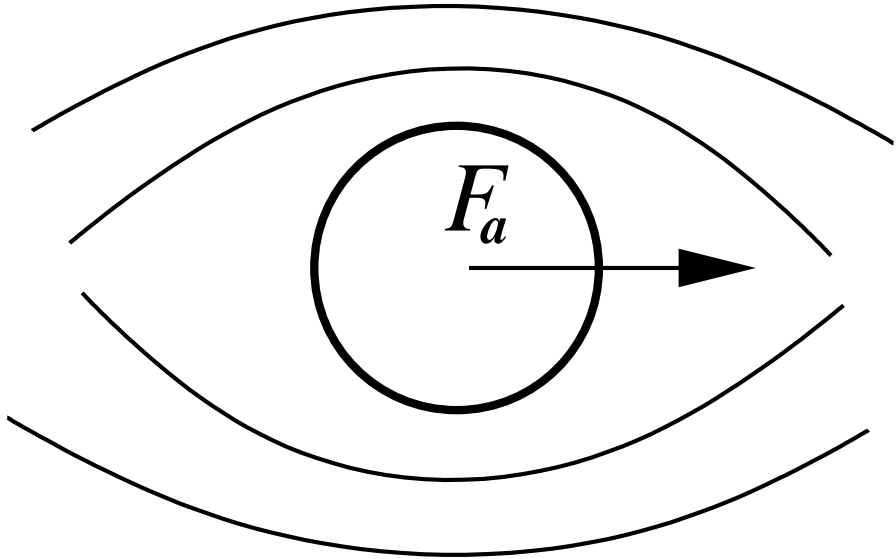,angle=0,width=12cm}}
\caption[]{\label{fig1} A body of mass $m_0$ and volume $V$ under the influence of
an applied force $F_a$. The body is surrounded by a static fluid whose density is $\rho$. }
\end{figure}

Consider a body  of mass $m_0$ and volume $V$ immersed in some
static fluid whose density is $\rho$ as shown in figure \ref {fig1}.
The body is initially at rest so that drag can be neglected. The dynamics of the situation shown in figure \ref{fig1} can be studied using Newton's second law
of motion \cite {symon}, $\sum {\bf F}_{\rm ext} = d {\bf p}/dt$, where the linear momentum is defined as
${\bf p}=m_0 {\bf v}$, with $m_0$ representing the mass and ${\bf v}$ the velocity. Now,   assuming that $m_0$ is constant, one can write the more restrictive relation $\sum {\bf F}_{\rm ext} = m_0 d{\bf v}/dt$ where $d{\bf v}/{dt}$ represents the acceleration, ${\bf a}$. Defined in this way, via Newton's second law of motion, $m_0$ represents the inertial mass, the characteristic that relates the sum of external forces on the body to the resulting acceleration. This inertial mass is indistinguishable from the gravitational mass,  which is defined in terms of the gravitational force it produces.
In this paper, this definition of mass will be sufficient for our purposes as a working concept but  readers seeking to get more insight in the concept of mass will find the recent work by  Roche \cite {roche} very illuminating.

Note, from figure \ref{fig1} that an applied force $F_a$ along the horizontal axis ($x$) will accelerate not only the body but also part of the fluid, with mass $S$, corresponding to the body's volume \cite {fluids}. Let us suppose that buoyance equals the body's weight so there is no force along $y$. Then, Newton's second law of motion  yields

\begin{equation}
F_a = m_0 a + S a = m^* a \,\,,
\label{added}
\end{equation}
where we have defined $m^* = m_0 + S$. The quantity
$S = \kappa \rho V$ is called the added mass while  the dimensionless parameter $\kappa$ relates to the body's geometry \cite {fluids} (e.g., $\kappa = 0.5$ for a sphere and $\kappa = 0.2$ for an ellipsoid moving ``end on" with the major axis twice the minor axis). The added mass term will become more familiar by recalling, for example, that it appears  easier to accelerate our hands in the air than in a bathtube.

From the experimental point of view, a numerical value for $m_0$ can be
obtained  by measuring
$a$ using, as inputs, the  values for $F_a$, $\rho$, $V$ and $\kappa$. In accordance with our supposition that $m_0$ is constant, the type of experiment displayed in figure \ref {fig1} will always generate the  same numerical value for $m_0$ irrespective of the input parameter values. Therefore, $m_0$ is also the  value which would be measured in the ideal situation  where the body is in the vacuum ($\rho=0$).  Let us call the constant $m_0$ the
bare mass. At the same time, the presence of the fluid can be treated
as an effective increase in the mass of the body for the purpose of determining the acceleration and therefore we can call $m^*$ the effective mass.
Eq. (\ref {added}) explicitly shows the main difference between bare and effective masses. Namely, when an interaction is present, the value $m_0$ changes to the value $m^*=m_0 + S$, where $S$ represents interactions. In this case one can also say that the bare mass has been dressed up by the interactions between the body and the
surrounding static fluid.

To illustrate the role of $m^*$ in classical physics let us consider a spherical balloon whose volume is $V$ filled with a gas with density $\rho_{\rm gas}=0.1 \rho$  where $\rho$ is the  air's density. The balloon is initially at rest so that the forces acting on it are its
weight , $W_{\rm balloon} = \rho_{gas} V g$ (neglecting the mass of the material of which the balloon is made), and the buoyant force, which corresponds to the weight of the displaced air, $W_{\rm air}= \rho Vg$. In this case, the effective mass is $m^*= m_0 + 0.5 V \rho$
where $m_0 = \rho_{\rm gas} V$. All forces act along the vertical axis and Newton's second law reads
\begin{equation}
W_{\rm air} - W_{\rm balloon}  = m^* a \,\,.
\end{equation}
An easy computation shows that the upward acceleration is given  by $a = 1.5 \, g$ which looks more reasonable than the value $a = 9 \, g$ obtained by neglecting
the added mass term, $S$. This illustrates the fact that the effective mass, which takes into account interactios with the medium, produces more realistic results.

\begin{figure}[htb]
\vspace{0.5cm}
\centerline{\epsfig{figure=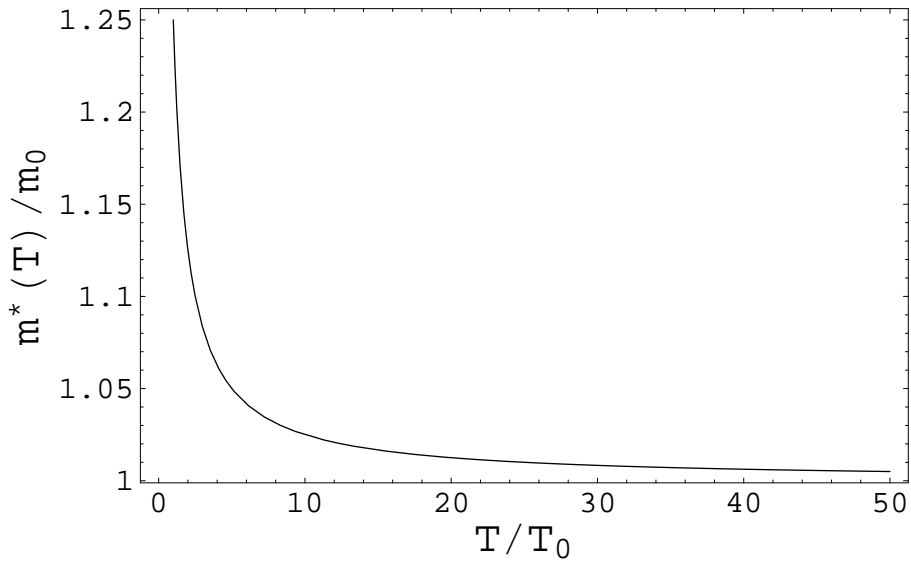,angle=0,width=12cm}}
\caption[]{\label{fig4} The dimensionless ratio $m^*(T)/m_0$ plotted as a function of $T/T_0$ taking $\rho_{\rm body} = 2 \rho_0$ and $\kappa=0.5$ (sphere). }
\end{figure}

Now, let us see how the temperature can change the effective mass of the immersed body so that analogies with modern theories can be further explored. Eq. (\ref{added}) clearly shows that $m^*$ increases linearly with the fluid's density, $\rho$. However, the temperature ($T$) also affects the added mass $S= \kappa V \rho$ since $\kappa$, $V$ and $\rho$ can change with $T$. For the sake of simplicity one can suppose that the body's shape and volume remain constant for a certain range of temperatures while the fluid's density changes. For example, consider the fluid to behave as an ideal gas whose equation of state is $\rho=P/(R T)$ where $P$ is the pressure and $R$ is the gas constant.
Considering the case where the pressure is constant the equation of state can be written as
$\rho = \rho_0 T_0/T$ where $\rho_0$ is a reference value at $T=T_0$. Then,
using $V= m_0 /\rho_{\rm body}$ one can write
\begin{equation}
m^*(T) = m_0 \left [1 + \frac {\kappa \rho_0}{ \rho_{\rm body}} \left ( \frac {T_0}{T} \right) \right ] \,\,.
\end{equation}
For numerical purposes, let us set $\kappa=0.5$ (sphere) and $\rho_{\rm body} = 2 \rho_0$.
Figure \ref {fig4} shows how $m^*(T)$ decreases with the temperature. This simple example will help us to understand, in section III,  a similar phenomenon, known as the chiral transition, which arises in  QCD where the effective mass of the quarks  decreases with the temperature \cite {kapusta}.

\subsection{ Sliding block in the presence of friction}

\begin{figure}[htb]
\vspace{0.5cm}
\centerline{\epsfig{figure=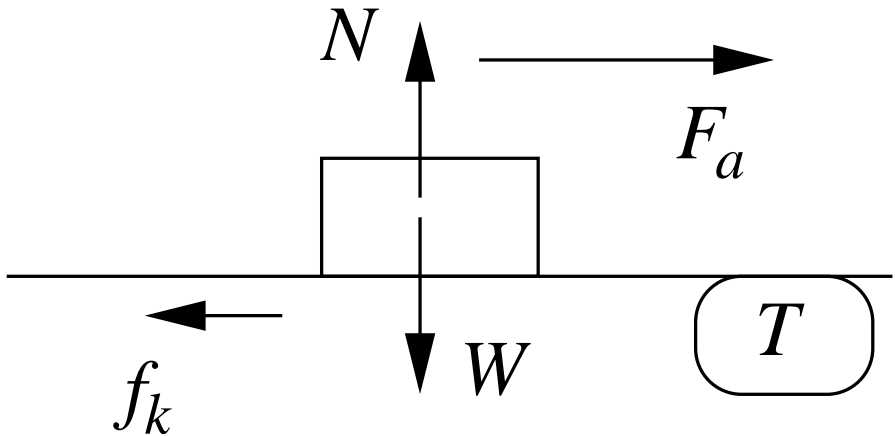,angle=0,width=12cm}}
\caption[]{\label{fig2} The usual diagrammatic representation of a sliding block, of mass $M_0$, in the presence of friction. }
\end{figure}

\begin{figure}[htb]
\vspace{0.5cm}
\centerline{\epsfig{figure=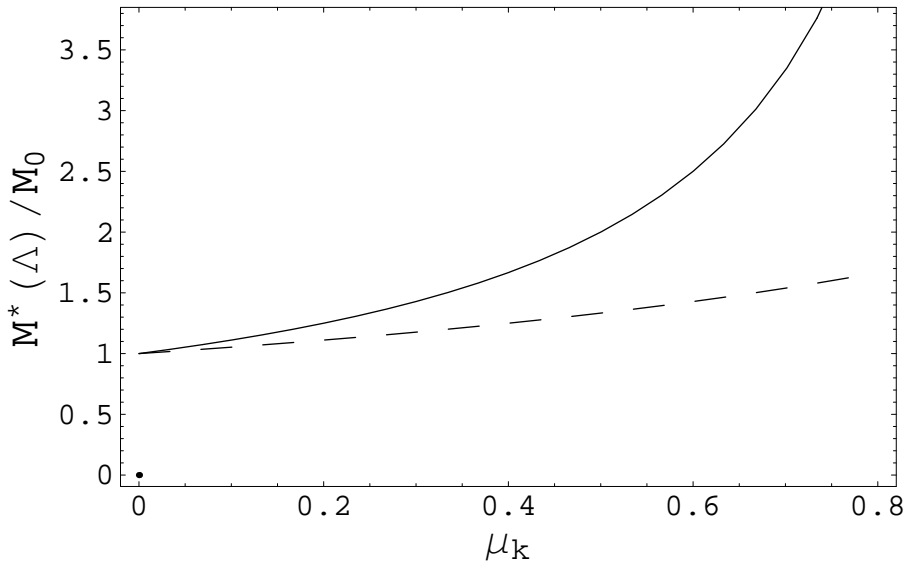,angle=0,width=12cm}}
\caption[]{\label{fig3} The dimensionless ratio $M^*(\Lambda)/M_0$ plotted as a function of $\mu_k$ for $\Lambda=1$ (continuous line) and $\Lambda = 2$ (dashed line).}
\end{figure}

Let us now consider a block of mass $M_0$, sliding over a flat surface, in the presence of friction and absence of air drag, as shown in figure \ref {fig2}. The forces acting on the body are its weight, ${\bf W}$, the normal force, ${\bf N}$, an applied force ${\bf F}_a$ and a kinetic frictional force, ${\bf f}_k$.  For the sake of the discussion to be carried out at the end of this subsection, the figure also shows a thermometer which records the temperature of the surface.
The forces can be decomposed in terms of $x$ and $y$ components by choosing $x$  parallel to the surface. Since there is no acceleration along $y$, one can write $N= W  = M_0 g$. Then, along $x$, one gets
\begin{equation}
F_a - f_k  = M_0 a \,\,,
\label{1}
\end{equation}
where $f_k = \mu_k |{\bf N}| = \mu_k M_0 g $. The dimensionless quantity $\mu_k$ represents the coefficient of kinetic friction between the block and the surface.
If we express the applied force in terms of the body's weight as $F_a = \Lambda M_0 g$, where $\Lambda$ is a dimensionless quantity, Eq. (\ref{1}) can be written as
\begin{equation}
\Lambda M_0 g - \mu_k M_0 g  = M_0 a \,\,,
\label{2}
\end{equation}
which shows that to have non null acceleration to the right one must have $\Lambda > \mu_k$.
For our purposes it is useful to rewrite Eq. (\ref {2}), in a more economical way, as
\begin{equation}
F_a =\Lambda M_0 g  = M^* a \,\,\,,
\end{equation}
where
\begin{equation}
M^*(\Lambda) = M_0 \left [ 1 - \frac {  \mu_k}{\Lambda} \right ]^{-1} \,\,\,.
\label{meff}
\end{equation}
At this stage some important remarks are in order. Comparing Eqs. (\ref {added}) and (\ref{meff}) we see that the latter depends on the intensity of applied force given by the dimensionless parameter $\Lambda$ so
that, unlike $m^*$, $M^*(\Lambda)$ is not a universal quantity (see figure \ref {fig3})
and cannot be treated as a true effective mass.
 Nevertheless, due to its mathematical form, this quantity will be useful to illustrate some issues related to perturbation theory since one can perform an expansion in powers of the ``coupling" parameter $\mu_k$ upon which an eventual temperature dependence may be imposed.
Then, a first glance at  other aspects of modern theories such as the importance of non perturbative methods and the renormalization group equations becomes possible, as we shall see in the next two sections. In particular, by exploiting the effects of temperature on the scale dependent equation (\ref {meff})  one  is able to present a simple illustration regarding the use  of perturbative/non perturbative techniques in modern theories. With this aim let us recall that in most classical mechanics courses $\mu_k$ is usually treated as a constant \cite {symon}. However, in practice, this
quantity can change under the influence of other parameters such as the temperature.
For example, a skier transfers energy to the snow underneath the skis,
thereby melting the snow and creating a thin film of liquid that reduces friction \cite {snow}.

\begin{figure}[htb]
\vspace{0.5cm}
\centerline{\epsfig{figure=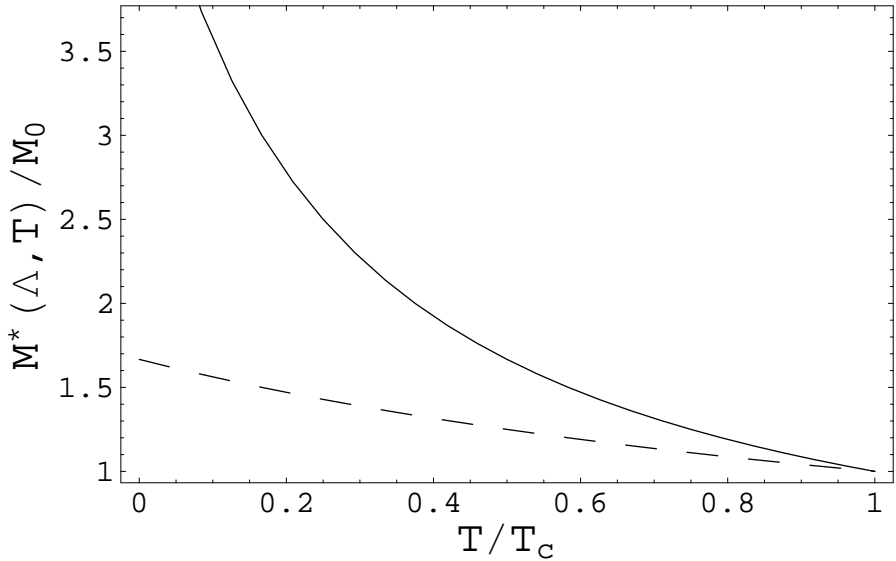,angle=0,width=12cm}}
\caption[]{\label{fig5} The dimensionless ratio $M^*(T,\Lambda)/M_0$ plotted as a function of $T/T_c$ for $\mu_k(0)=0.8$. The solid line represents the $\Lambda=1$ case while for the dashed line the value is $\Lambda=2$. }
\end{figure}

In other physical systems \cite {atrito}, heat-activated plastic deformations and other thermal effects can also induce a temperature dependence on $\mu_k$.
In this section, we will suppose  that our $\mu_k$
also changes with the  temperature.
Figure \ref{fig2} shows the thermometer of some device which generates these thermally induced changes  (the exact manner in which this happens does not concern us here). Let us just suppose that the experiments show a decrease of $\mu_k$, with increasing temperatures, which can be mathematically described by
\begin{equation}
\mu_k (T) = \mu_k(0) \left [ 1 - \left (\frac {T}{T_c} \right )\right ] \,\,\,,
\label {mut}
\end{equation}
where $T \le T_c$ so that $0 \le \mu_k(T) \le \mu_k(0)$ (one could also have multiplied Eq. (\ref {mut}) by the step function, $\theta(T_c-T)$). Now, plugging this equation into Eq. (\ref {meff}) one can see how the effective mass ``runs" with  the temperature. Setting  $\mu_k(0)=0.8$ one obtains figure \ref{fig5}, which  displays $M^*(T,\Lambda)/M_0$,  signaling that $M^*(T_c,\Lambda)=M_0$.
To conclude this section let us remark that in both situations discussed above
the numerical values of the parameters  $\rho$ (subsection II.A) and $\mu_k$ (this subsection) give the intensity of the relevant interactions. To reinforce, once again, further analogies between the classical and modern descriptions of the effective mass note that  within the latter such quantities are called the coupling constants. As discussed in the next section, the one which
is relevant for QCD ($\alpha_s$) decreases
 with increasing temperatures (asymptotic freedom) somewhat like our $\rho(T)$, and $\mu_k(T)$ inducing the effective  quark mass to approach its bare value.

\section{Bare and effective mass in quantum field theories}

The concepts of bare and effective mass are widely employed in modern
quantum field theories (QFT). Within these theories one usually starts from a Lagrangian
density written in terms of the bare masses and interactions whose intensity is
given by coupling parameters, $\alpha$. As in the classical case the interactions induce a change $S_q$ (where $q$ stands for quantum) in the value of the observed (effective) mass, $m_{0,q} \to m^*_q=
m_{0,q} + S_q$. In principle, one must evaluate the quantity $S_q$, which is called the self energy. In classical physics the self energy of a system is the contribution to its energy  resulting from the interaction between different parts of the system. In the many body problem in quantum mechanics the self energy is the difference between the
energy of the quasi particle associated with the particle and the energy of the bare particle. The particle interacts with its surrounding medium, which in turn acts back on the original particle. In QFT the self energy of a particle is the contribution to the energy (mass) of the particle due to the virtual emission and absorption of particles.

The evaluation of $S_q$, within QFT, is far from being a trivial matter. One reason for this is due to the fact that apart from the obvious $\alpha$ dependence the self energy often depends on the particle energy-momentum, $p$, which has to be integrated over. At the same time, if the particle is in contact with a thermal bath and/or in a dense medium the self energy will also depend on the temperature ($T$) and/or on the density ($\rho$). In a  typical QFT computation of the self energy one considers equations like
\begin{equation}
S_q = \alpha L_1(p) + \alpha^2 [L_1^2(p) + L_2 (p)] +...+\alpha^n [L_1^n(p)+ L_1^{n-1} L_2(p)+...+ L_n (p)]+...
\label{sigma}
\end{equation}
where $L_i(p)$ represents integrals over $p$.
Most of the time,
the exact evaluation of $S_q$ is not possible and one
must employ some approximation technique. For example, if $\alpha << 1$ one can use perturbation theory by evaluating just the first few terms on the RHS of Eq. (\ref {sigma}). However, if $\alpha \ge 1$ non perturbative techniques must be employed. In this type of approximation one re-sums an infinite subset of contributions. For example, if for some physical reason  the integrals $L_1(p)$ appear to be more important than the others it becomes possible to write
$S_q$ as the series
\begin{equation}
S_q = \alpha L_1(p) + \alpha^2 L_1^2(p)  +...+\alpha^n L_1^n(p)+...= \frac {\alpha L_1(p)}{\left [ 1 - \alpha L_1(p)\right ] } \,,
\label{hartree}
\end{equation}
which is non perturbative in the coupling $\alpha$.
However, in many cases a major problem arises when the integrals over $p$ produce divergences since the  measurable
 $m^*_q$ is finite. Let us illustrate the above discussion with a few examples.

\subsection{The neutron effective mass at high densities}

A nice illustration is provided by the neutron effective mass at high densities which is relevant for studies concerned with neutron stars \cite {kapusta}. At
zero density the neutron has a (bare) mass $m_{0,q}=939.56 \, {\rm MeV}/c^2$, where $c$ is the speed of light. However, within the confines of a neutron star this quantity changes drastically as a result of interactions which, as discussed, are taken into account by the self energy. In this evaluation one is not concerned by the temperature since this quantity is approximately zero in a neutron star. Typically, the models employed in this computation consider $\alpha > 1$ and a resummation technique like Eq. (\ref {hartree}) must be employed. Finally, when one is only interested in density effects the integrals are carried up to a finite (Fermi) momentum value without producing any divergence. The result of this evaluation shows that the effective neutron mass $m^*_{q} = m_{0,q} + S_{q}(\alpha, \rho)$ decreases with density \cite {kapusta}.

\subsection{Quantum electrodynamics}

In order to discuss the renormalization problem let us now
consider quantum electrodynamics (QED) at zero temperature and density.
For this theory, $\alpha = 1/137$ and perturbation theory works as confirmed by  many
successful predictions.  In the introduction it was mentioned how
 the values which represent the electron mass inside ($m^*_{q}$) and outside ($m_{0,q}$)  a solid are different. The theoretical solid state physicist may predict both values by evaluating the (always finite) self energies whereas his/her experimentalist colleague is able to measure both quantities. The theoretical situation is rather different at the fundamental (QED) level since $S$ diverges because the integrals $L_i(p)$ are evaluated over all possible (even infinite) momentum values which concern the virtual particles.
 As with the immersed body, the electron inside a solid, or the neutron in a dense medium,  the interactions change the mass value but this time to an unacceptable infinite value.
However, one should note a major difference concerning these four cases. Namely, in the first three situations one can always know, experimentally, the bare  mass values by turning off the interactions.  Note carefully that what the solid state physicist calls the electron's bare mass is what the high energy physicist calls the effective mass.
This is because both physicists are interested in different energy scales and the most relevant interactions for the first are the ones between the electron and the lattice.  In principle, the solid state physicist can measure how an electron inside the solid responds to an externally applied force. At the same time, she/he can  measure how an identical electron outside the solid responds to an applied force thereby determining what he/she calls the bare mass. The last procedure  would correspond to ``turning the interaction off". However, in a fundamental field theory like QED this is not possible because the interactions are inherent \footnote {Technically this happens because fundamental theories are built from
a gauge invariant principle which is violated if there are no interactions between the charges  and gauge bosons (photons, for QED).} so that for the high energy physicist who is concerned with the interactions between the electrons and virtual photons the interaction is always ``on". In other words, the electron mass is always dressed up by the interaction with virtual particles and it can, eventually, be further dressed up by other type of  interactions like the one due to the presence of a lattice.
Therefore, at the most fundamental level, the real bare mass of free electrons can never be measured. Within the renormalization programme one uses this fact to redefine
 the bare mass so that it exactly cancels the divergence produced by the self energy producing a finite value for the observable $m^*_q$. This procedure brings in some arbitrariness \cite {cheng} which will be briefly addressed in the last section.

\subsection{Quantum chromodynamics}

In the next two sections we will discuss some analogies between  our classical systems
and the fundamental theory of strong interactions, quantum chromodynamics (QCD).
To make this discussion more profitable let us review some relevant QCD issues.
The fundamental particles involved in this theory are six different types of quarks whose respective masses appear with two different values in some particle data tables \cite {pdb}. One value refers to the  current mass ($m_c$) and the other to the  constituent (or effective) mass value, $m^*_q$. For example, the up quark ($u$), which is the lightest, has a current mass close to $4 \,\, {\rm MeV}/c^2$ while
its constituent mass is about
$363 \,\, {\rm MeV}/c^2$ in hadrons and $310 \,\, {\rm MeV}/c^2$ in mesons. Quarks, which are
fermions, interact among themselves by the exchange of spin-1 bosons called gluons.
Contrary to photons, gluons interact with each other producing an
effect known as asymptotic freedom which basically means that the QCD coupling ($\alpha_s$) decreases with high energies (temperatures) or small distances.
Asymptotic freedom implies that there should be a qualitative change in the properties of hadronic matter as the temperature and/or the density is increased. A cold and dilute system could be described in terms of pions, nucleons and other hadrons. However, in a very hot and/or dense system quarks and gluons would be free to roam in what is called a quark-gluon plasma. One can envisage that a phase transition from hadron matter to quark-gluon plasma could have occurred in the early universe. As far as the actual computations are concerned asymptotic freedom means that one must revert to non perturbative techniques when considering low
temperatures since $\alpha_s$ is large in those regimes. On the other hand one can, at least in principle, use perturbation theory when performing high temperature evaluations. For our purposes it is useful to consider the evaluation of the quark thermal effective mass, $m^*_q(T) = m_{0,q} + [\alpha_s (T) L_1(p)+...]$. As within QED the
integrals  usually produce a  divergent ($D_1$) plus a finite piece ($F_1$),
$L_1(p) = D_1 + F_1$. One possibility, to obtain renormalized results, is to define the bare mass as $m_{0,q} = m_c - \alpha_s(T) D_1$ where $m_c$ represents the finite current mass. One then obtains the finite constituent thermal mass, $m^*_q(T) = m_c + \alpha_s(T) F_1 +...$. The decreasing of $\alpha_s(T)$ with increasing temperatures shows that $m^* \to m_c$ at temperatures close to $10^{12} {\rm K}$. Since a quantitative discussion of these issues within QCD is too difficult to be carried out here one should refer to the classical examples, previously discussed, in order to get some mathematical insight.

A few remarks are now in order. First, note that the numerical values of the quark masses found in the literature depend on the way they are defined (model dependent). Also, since individual quarks cannot be observed  their masses are not  directly observable. Finally, note that many times in the literature the terms current mass and bare mass are taken to represent the same quantity. This language problem arises because there are different ways of implementing the renormalization programme \cite {cheng} (e.g., counterterms or multiplicative renormalization constants). To avoid confusion and to pursue our analogies we shall use, in the next section, the term bare mass as representing the current mass since renormalization issues will not appear.

\section{Writing a diagrammatic equation for the effective mass}

\begin{figure}[htb]
\vspace{0.5cm}
\centerline{\epsfig{figure=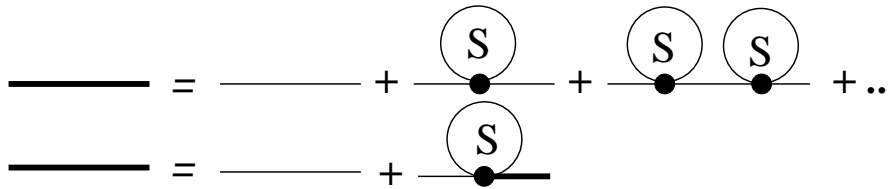,angle=0,width=12cm}}
\caption[]{\label{fig6} A diagrammatic expression for  $M^*(\Lambda)$, represented by the thick line. The bare mass, $M_0$, is represented by the thin straight line while the small black dot represents $\mu_k$ and the bubble represents $S(\Lambda)$.}
\end{figure}

Using Eq. (\ref {meff}) one can also give a flavor of Feynman's diagrammatic way of doing physics. To use a notation similar to the one employed in QFT  let us  define the quantity $S(\Lambda) =  1/(M_0 \Lambda)$ with units of ${\rm kg}^{-1}$. Then, Eq. (\ref {meff}) can be written as
\begin{equation}
M^*(\Lambda) = M_0 \left [ 1-  \mu_k M_0 S(\Lambda) \right ]^{-1} \,\,,
\label{whole}
\end{equation}
and, further, expanded in powers of $\mu_k$ yielding
\begin{equation}
M^*(\Lambda)= M_0 + M_0[\mu_k S(\Lambda)]M_0 + M_0[\mu_k S(\Lambda)]M_0[\mu_k S(\Lambda)]M_0 + \dots
\label{serie}
\end{equation}
The above equation can be rewritten in a self-consistent way as
\begin{equation}
M^*(\Lambda)= M_0 + M_0[\mu_k S(\Lambda)]M^*(\Lambda) \,\,\,.
\label{self}
\end{equation}
One can represent both Eqs. (\ref {serie}) and (\ref {self}) in a diagrammatic way by
attributing some rules to each of the quantities, $M^*(\Lambda)$, $M_0$,$\mu_k$ and $S$. Let us represent the first by a thick line, the second by a thin line, the third by a  small black dot and $S$ by a bubble to get
figure \ref{fig6}.
In modern physics, such as QFT, it is often impossible to know the whole series representing a physical quantity such as   Eq. (\ref {whole}). Instead one knows, for each theory, a set of (Feynman) mathematical rules to compute the graphs shown
in figure \ref{fig6}. However, the bubbles can also have an inner structure which can be very complicated.  In fact, in QFT the bubbles are the diagrammatical representation of the
integrals $L_i(p)$ displayed in Eq. (\ref {sigma}). As already emphasized, this calls for approximations in the computation of quantities like the effective mass. Let us use our results for
the sliding block  supposing that
 $\mu_k << 1$ and $\Lambda \sim 1$. Then, Eq. (\ref {serie}) can be written as
\begin{equation}
\frac{M^*(\Lambda)}{M_0} \simeq 1 + \frac{\mu_k}{\Lambda}  \,,
\label{first}
\end{equation}
making a good prediction of the actual $M^*(\Lambda)/M_0$ numerical value. However, as $\mu_k$ increases more terms are needed if one wants more precision. For example, setting $\Lambda=1$, the exact result for $\mu_k=0.01$ is $M^*(1)=1.{\overline {01}} \, M_0$ while our perturbative first order prediction is $M^*(1)= 1.01\, M_0$. But if $\mu_k=0.8$ the exact result is  $M^*(1)=5 \, M_0$ while the first order perturbative result  performs rather poorly, predicting
$M^*(1)=1.8 \, M_0$.  We therefore conclude that the failure or success of the lowest orders perturbative predictions depend on the actual value of the parameter $\mu_k$ unless $\Lambda >> \mu_k$. As we have seen in the previous section, the value of $\mu_k$ can  change with the temperature (see Eq. (\ref{mut})). Then, by replacing $\mu_k \to \mu_k(T)$ in Eq. (\ref {first}) while remaining within the limit $\Lambda \sim 1$ one sees that, like in QCD, lowest order approximations will perform better at higher temperatures than at lower temperatures. Finally, it is worth remarking that the development of reliable non perturbative techniques to tackle QCD is a very important domain of modern research. The reader interested in a formal discussion of perturbation theory via Feynman diagrams in classical mechanics is refered to a recent work by Penco and Mauro \cite {penco}.

\section{A comment on scale invariance}
As we have emphasized there is a major difference between our two classical quantities $m^*$ and $M^*(\Lambda)$ given by Eqs. (\ref {added}) and (\ref {meff}) respectively. Namely, the first is a true effective mass being independent of the intensity of applied force  ($\Lambda$) ultimately used to measure it. Our $M^*(\Lambda)$ on the other hand clearly displays the fact that Eq. (\ref {meff}) was obtained by some redefinition of the terms entering Newton's second law for the sliding block. Although this equation has no advantages concerning practical applications of Newton's second law one can again profit from its form to illustrate QFT related issues. First, note that the quantity $M^*(\Lambda)$ would be $\Lambda$ independent only if $\mu_k$ itself had a compensating $\Lambda$ (or $F_a$) dependence, that is if $\mu_k = \mu_k(\Lambda)$. However, this scenario cannot be considered simply because experimental observation rules it out. On the other hand,
in QFT calculations one often ends up with quantities which depend upon some arbitrary energy scale, $\Lambda_q$, which is introduced during the formal mathematical manipulation of the  integrals, $L_i(p)$, in a procedure known as regularization \cite {ren,hans}. As we have mentioned in section III.B many of these integrals diverge in the limit $p \to \infty$ (ultra violet divergence) and one way of regularizing  the theory requires the introduction of scale parameters ($\Lambda_q$) which, {\it contrary} to ($\Lambda$), do not belong to the original theory. So, in QFT one has $m^*_q(\Lambda_q)$ but needs to further  require that the predicted (and measured) value for this quantity be $\Lambda_q$ independent. Mathematically,  this requirement translates into
\begin{equation}
\frac {d m^*_q(\Lambda_q)}{d\Lambda_q} = \frac {\partial m^*_q(\Lambda_q)}{\partial \Lambda_q} + \frac {\partial \alpha}{\partial \Lambda_q}\frac {\partial m^*_q(\Lambda_q)}{\partial \alpha}
+ \frac {\partial m_0}{\partial \Lambda_q}\frac {\partial m^*_q(\Lambda_q)}{\partial m_0}=0 \,\,.
\label{rg}
\end{equation}
which is a renormalization group type of equation \cite {cheng}. We see that
the coupling constant $\alpha$ is allowed to run with $\Lambda_q$, that is $\alpha = \alpha(\Lambda_q)$ in a way  dictated by the term $\partial \alpha/\partial \Lambda_q$ which is called the
$\beta$-function. For QCD the $\beta$-function is such that the coupling constant decreases with increasing
energies (or temperatures) leading to asymptotic freedom, a fact confirmed by experiments.

\acknowledgments

This work has been  partially supported by Conselho Nacional de
Desenvolvimento Cient\'{\i}fico e Tecnol\'{o}gico (CNPq-Brazil).
I thank  Hugh Jones for his useful comments and  especially for suggesting the inclusion of Sec. II.A. I dedicate this work to the the memory of my father, who valued education so much.

\eject

\centerline {FIGURE CAPTIONS}

{ FIG 1: A body of mass $m_0$ and volume $V$ under the influence of
an applied force $F_a$. The body is surrounded by a static fluid whose density is $\rho$. }

{Fig 2: The dimensionless ratio $m^*(T)/m_0$ plotted as a function of $T/T_0$ taking $\rho_{\rm body} = 2 \rho_0$ and $\kappa=0.5$ (sphere). }

{Fig 3: The usual diagrammatic representation of a sliding block, of mass $M_0$, in the presence of friction. }

{Fig 4: The dimensionless ratio $M^*(\Lambda)/M_0$ plotted as a function of $\mu_k$ for $\Lambda=1$ (continuous line) and $\Lambda = 2$ (dashed line).}

{Fig 5: The dimensionless ratio $M^*(T,\Lambda)/M_0$ plotted as a function of $T/T_c$ for $\mu_k(0)=0.8$. The solid line represents the $\Lambda=1$ case while for the dashed line the value is $\Lambda=2$. }

{Fig 6: A diagrammatic expression for  $M^*(\Lambda)$, represented by the thick line. The bare mass, $M_0$, is represented by the thin straight line while the small black dot represents $\mu_k$ and the bubble represents $S(\Lambda)$.}
\end{document}